\documentclass[3p]{elsarticle}

\usepackage{latexsym}
\usepackage{epsfig}
\usepackage{amssymb}
\usepackage{color}
\newcommand{\be}{\begin{equation}}
\newcommand{\ee}{\end{equation}}
\newcommand{\ba}{\begin{eqnarray}}
\newcommand{\ea}{\end{eqnarray}}
\newcommand{\lsim}   {\mathrel{\mathop{\kern 0pt \rlap
  {\raise.2ex\hbox{$<$}}}
  \lower.9ex\hbox{\kern-.190em $\sim$}}}
\newcommand{\gsim}   {\mathrel{\mathop{\kern 0pt \rlap
  {\raise.2ex\hbox{$>$}}}
  \lower.9ex\hbox{\kern-.190em $\sim$}}}

\begin{document}
\title{The Continuous Tower of Scalar Fields as a System of Interacting Dark Matter - Dark Energy}
\author{ Paulo Santos$^\ddag$}
\ead{p.g.d.santos@astro.uio.no}
\address{$\;$\\

$^\ddag$Institute of Theoretical Astrophysics, University of Oslo, P.O. Box 1029 Blindern, N-0315 Oslo, Norway}

\date{\today}

\begin{abstract}
This paper aims to introduce a new parameterisation for the coupling $Q$ in interacting dark matter and dark energy models by connecting said models with the Continuous Tower of Scalar Fields model.
Based upon the existence of a dark matter and a dark energy sectors in the Continuous Tower of Scalar Fields, a simplification is considered for the evolution of a single scalar field from the tower, validated in this paper.
This allows for the results obtained with the Continuous Tower of Scalar Fields model to match those of an interacting dark matter - dark energy system, considering that the energy transferred from one fluid to the other is given by the energy of the scalar fields that start oscillating at a given time, rather than considering that the energy transference depends on properties of the whole fluids that are interacting.

\end{abstract}


\maketitle

\section{Introduction}

One of the deepest problems in Cosmology lies in the lack of understanding of about $95\%$ of the energy content of the universe, the so called dark sector \cite{Planck}.

Composing this sector, two components with clearly distinct behaviours exist, dark matter (DM) and dark energy (DE), with the energy density of DM being around half the energy density of DE today.
While DM behaves like matter, lacking only the capability to interact electromagnetically, DE behaves like a cosmological constant, a fluid with constant energy density (details on why considering a simple cosmological constant would present some problems can be found in \cite{CP}).

Scalar fields are generally good candidates for both components of the dark sector, as it is possible to replicate the same behaviour of DM or DE when an adequate potential for the scalar field is chosen.
While most DE models consider scalar fields \cite{DE}, research on DM spans from Cosmology to Particle Physics.
On the cosmological side, there are several models considering from Axions \cite{Axions} to discrete towers of scalar fields \cite{TDM}, with the potential considered for the scalar fields in both mentioned cases being $V={1 \over 2}m^2\phi^2$.
Despite most models consider scalar fields to explain a single component of the dark sector, there are some models that aim to account for both dark components, like the Continuous Tower of Scalar Fields (CTSF) model presented in \cite{Tower}.

~

Another field of study that is particularly interesting when it comes to the dark sector lies in possible interactions between DM and DE, which is an obvious possibility when taking into account that the densities of both components are quite similar today, as mentioned before.

The fact that those energy densities are of the same order is the so-called "coincidence problem", despite many argue it is not really a problem (a good discussion on this topic can be found in \cite{Coin}).
This possible interaction between the two components of the dark sector could possibly alleviate this problem, and explain why the energy densities of both components are quite similar today. 

~

The most standard way to model this interaction consists in considering both DM and DE as fluids and add an interaction term to the continuity equations of both fluids in the following way:

\ba
\begin{array}{c l}      
    \dot{\rho}_{DE}+3H(1+w)\rho_{DE}=Q~,\\
    ~\\
    \dot{\rho}_{DM}+3H\rho_{DM}=-Q~,\\
\end{array}
\label{DDI}
\ea

The first parameterisations for the coupling term $Q$ were given by $Q=\delta_{DM}H\rho_{DM}$, \cite{Q}, which was later generalised to $Q=\delta_{DE}H\rho_{DE}+\delta_{DM}H\rho_{DM}$, \cite{QM}.
In said parameterisations, the $\delta$ terms give the strength of the coupling, which can either be constant or variable in time.
By taking into account that the ratio $R=\rho_{DE}/\rho_{DM}$ today must agree with the observations, it is possible to constraint $\delta$, which must obey the following relation for the models mentioned, $|\delta |<<1$.
Also, it is relevant to mention that, rather than considering a $\delta H$ factor, some models chose to describe the strength of the interaction by a $\Gamma$ factor instead.

With this formalism, it is necessary to resort to dynamical variables and a dynamical systems kind of analysis in order to understand how the energy densities of DE and DM evolve in time, given that it is not possible to solve the differential equations presented in \ref{DDI} independently.

It is also relevant to mention that DM and DE do not necessarily need to be fluids, and in fact  is also possible to consider the interaction between a fluid and a scalar field, the so-called interacting quintessence model \cite{IQM}, or even the interaction between two scalar fields \cite{QSF}, in which the interaction is achieved by considering a modified potential that contains the coupling between the two scalar fields.

~

In this paper, the parameterisation for the coupling $Q$ considered will be considerably different to the ones presented before.

Rather than looking at DM and DE as just fluids, those two components will be also looked as sets of scalar fields that mutate from a DE state into a DM state, and hence the parameterisation of $Q$ will describe this transference of scalar fields.
Rather than depending on the total energy of the DM and DE fluids, as done before, $Q$ will depend on the energy of the scalar fields that shift from one state into the other.

The physical motivation for such a parameterisation for $Q$ lies in the CTSF model, in which scalar fields start by behaving as DE, with said behaviour shifting into DM at later times. 

Therefore, the paper will develop as follows.
In section 2, a simplified CTSF model, considering a simplified behaviour for the single scalar fields that compose the CTSF is presented and validated, while in section 3, the interacting DM-DE equations are used in order to obtain the same results for the evolution of the CTSF as in section 2.
The paper ends with section 4, which features the final discussions of the results obtained in the paper.

\section{Simplified Continuous Tower of Scalar Fields}

As mentioned in the introduction, scalar fields are usually one of the best candidates to explain the dark sector, given that they can behave as DM or DE, depending on the potential chosen.

In \cite{Tower}, a new innovative model was proposed to unify the dark sector.
To obtain this unification, a set of minimally coupled scalar fields with $V={1 \over 2}m^2\phi^2$ and a continuous distribution of masses is considered.
Since no coupling is considered between the scalar fields in said set, each scalar field will independently obey to the usual Klein-Gordon equation:

\be
\ddot{\phi}_m+3H\dot{\phi}_m+m^2\phi_m=0
\label{kg}
\ee

\noindent and the energy density for each scalar field will therefore be given by:

\be
\rho_{\phi_m}={1\over2}\dot{\phi}_m^2+{1\over2}m^2\phi_m^2
\ee

With the potential mentioned before, the Klein-Gordon equation presented in eq.\ref{kg} is equivalent to the equation to a damped harmonic oscillator, with $\omega_0=m$.

~

Considering the previous analogy, it becomes clear that it is possible to choose a set of initial conditions for which the scalar fields are constant until $t \approx T$, oscillating afterwards.

Therefore, the scalar field will behave like DE until it starts oscillating.
Afterwards, the behaviour will depend on how $H$ evolves in time as well, and it will therefore depend on the era considered.

During radiation and matter domination, $H$ can be parameterised as $H=p/t$, with $p=1/2$ and $p=2/3$ respectively, which means the energy of the scalar field will then decay with $a^{-3} \propto t^{-3p}$, behaving like DM.

When the CTSF dominates, the situation becomes much more complex.
Generally, it is not possible to parameterise $H$ with a simple expression before solving the Klein-Gordon equation, due to the fact that $H$ will now depend on the energy densities of the individual scalar fields on the CTSF, which means those scalar will in fact be now coupled gravitationally.
This leads to two distinct cases for the behaviour of the individual scalar fields, which will depend on which kind of scalar fields that dominates.

If the CTSF is dominated by the scalar fields that are already oscillating, then $H$ can be parameterised by $H=p/t$, with $p=2/3$, just like during matter domination, and hence the evolution of the individual scalar fields will be the same as during said era, behaving like DM.
This case is equivalent to matter domination with DM being given by the sector of the CTSF that contains the oscillating scalar fields, and therefore the results presented for matter domination will be valid in this case as well.

On the other hand, if most of the energy of the CTSF is due to the scalar fields that are frozen, the behaviour of the individual scalar fields will be considerably different.
When $t \approx T$, the scalar fields will start experiencing over-damped oscillations for which $\dot{\phi}_m\approx0$.
This means that, despite the scalar fields are oscillating, their behaviour is actually closer to DE than to DM, despite it is still something in between those two regimes.
The longer this regime lasts, the longer the scalar fields experience over-damped oscillations, which means the behaviour of a single scalar field is hard to parameterise, but also that the transition from a DE state to a DM state cannot be approximated by an instantaneous transition, as done in the previous eras.

It is also relevant to mention that the asymptotic behaviour of the scalar fields is a DM behaviour, which means that if a given scalar field is already behaving like DM, its behaviour will not change when the universe enters an era dominated by the CTSF.

~

For the two simple cases mentioned before, radiation and matter domination domination, the total evolution of $\rho_{\phi_m}$ can therefore be approximately described by two separate cases.
If the scalar field starts oscillating during radiation domination, its evolution can be roughly given by:

\ba
\rho_{\phi_m}=\rho_{\phi_m}(t_{ini})\times\left\{
\begin{array}{c l}      
    1 & ~,~t<t_d\\
    ~\\
    (mt)^{-3/2} & ~,~t_d<t<t_e\\
    ~\\
    \sqrt{t_e}m^{-3/2}t^{-2} & ~,~t_e<t\\
\end{array}\right.
\label{EvoR}
\ea

\noindent with $t_d=1/m$, which approximates the time for which the scalar fields start oscillating, given that $\omega_0=m$ and $H$ is of order $1/t$, and $t_{ini}$, which is the initial time, while $t_e$ is the time of the transition from radiation to matter domination, which means it will be relevant for $m>1/t_e$.
If the scalar field has $m<1/t_e$, it will not start oscillating before matter domination and so its evolution can be approximately described by:

\ba
\rho_{\phi_m}=\rho_{\phi_m}(t_{ini})\times\left\{
\begin{array}{c l}      
    1 & ~,~t<t_d\\
    ~\\
    (mt)^{-2} & ~,~t_d<t\\
\end{array}\right.
\label{EvoM}
\ea

Physically, this can be seen as if the scalar fields have only the possibility to be in two different states, a DE one (when the scalar fields are frozen), and a DM one (when the scalar fields oscillate), with the transition from one state to the other being instantaneous at $t=t_d$.

However, due to the reasons mentioned before, this kind of simplification is only valid until the DE sector of the CTSF starts to dominate.

~

For the evolutions of the scalar fields, described by eqs.\ref{EvoR} and \ref{EvoM}, to happen, the following initial conditions must be considered:

\ba
\begin{array}{c l}      
    \phi_m(t_{ini})=A\left\vert\frac{\sigma}{m}\right\vert^b~,\\
    ~\\
    \dot{\phi}_m(t_{ini})=0~,\\
\end{array}
\ea

\noindent as presented in \cite{Tower}. With said initial conditions, the initial energy density for the scalar field is given by:

\be
\rho_{\phi_m}(t_{ini})=Cm^{2(1-b)}
\label{inen}
\ee

\noindent with $C={1 \over 2}A^2\sigma^{2b}$, since the potential term in the energy dominates.

In the previous expressions, $A$ is the amplitude for the scalar field, $\sigma$ is the width of the distribution for the CTSF (which can be roughly translated into the highest mass considered in the CTSF), $m$ the mass of a scalar field in the CTSF and $b$ a parameter that allows the energy density to be shaped over the mass ($b=1$ allows for the energy of the scalar fields to be independent of the mass while for $b<1$ the energy grows with the mass and for $b>1$ the opposite occurs).

It is relevant to mention that the previously presented expressions for the initial conditions, originally presented in \cite{Tower}, are quite general given said paper aimed to study the implications of the CTSF without paying special attention to the underlying mechanism generating it. This resulted in a more general model, but it is still a model with proper physical motivation, given that the CTSF can in fact be a good approximation to a discrete tower with a large number of scalar fields and a small mass gap. Therefore, unparticle physics \cite{Unparticles}, braneworld scenarios (like the Randall-Sundrum \cite{RS} or DGP \cite{DGP} models), or even more exotic setups could provide the underlying mechanism to generate the CTSF (a deeper discussion on the subject is presented in the introduction of \cite{Tower}).

~

\begin{figure}[ht]
\includegraphics[width=8cm]{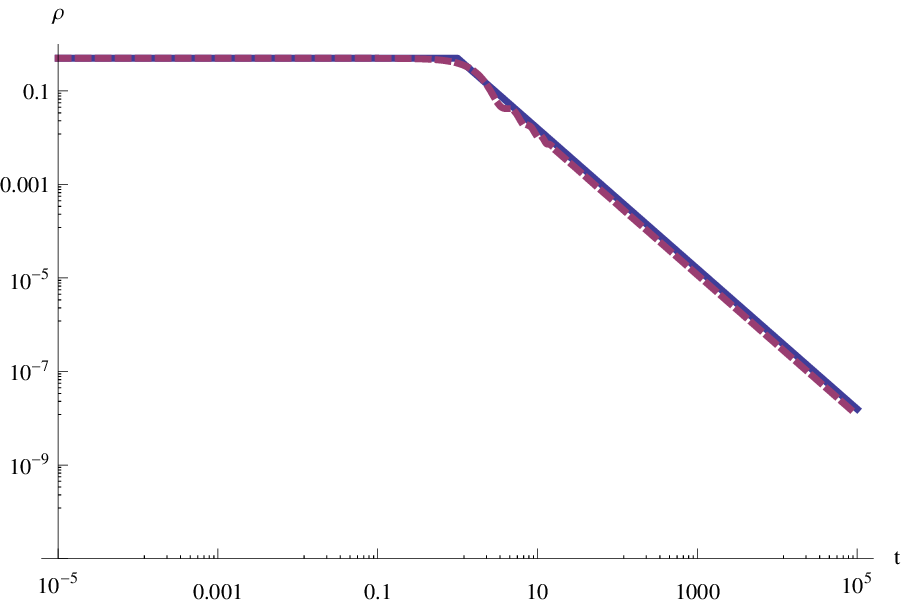}
\includegraphics[width=8cm]{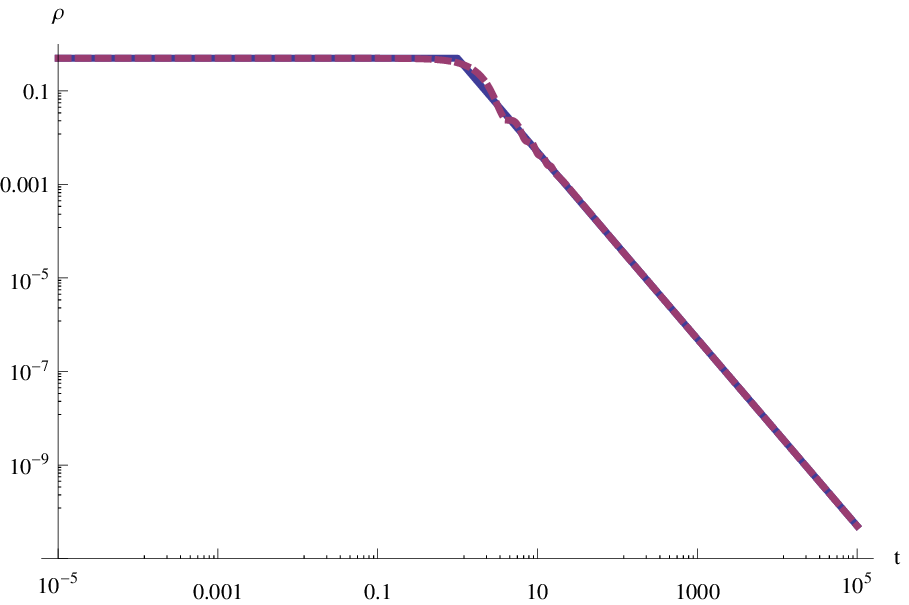}
\caption{
In the left panel, $\rho_{\phi_m}$ is plotted during radiation domination, while the right plot shows the same quantity during matter domination. 
In both plots, the thick line denotes the approximation considered in this section, while the dashed lined refers to the analytic solution obtained in \cite{Tower}.
In both cases, the energy density decays like matter, with $a^{-3}$.
}
\label{1Field}
\end{figure}

By plotting the simplified solutions, presented in eqs.\ref{EvoR} and \ref{EvoM}, together with the analytic solutions, obtained in \cite{Tower}, it is possible to notice that the only significant differences in behaviour occur around $t=1/m$, as expected.
However, afterwards the value of $\rho_{\phi_m}$ is slightly shifted during radiation domination, which does not occur for matter domination, as shown in figure \ref{1Field}.

Given said solutions provide good approximations to the analytic solutions, it is now possible to use them to obtain a new simplified approximation solutions for the energy density of the whole CTSF, $\rho_T$, during radiation and matter domination.

However, it is relevant to point out another simplified solution for $\rho_T$, during radiation and matter domination, was already presented in \cite{Tower}, but as will be seen in the next section, the new set of solutions obtained afterwards will be crucial to show the connection between the CTSF and models with interacting DM-DE, as this paper aims to.

The process to determine this new simplified solution for $\rho_T$ is similar to what was done in \cite{Tower}, but to keep the calculations very simple, further simplifications to the CTSF model were considered.

As mentioned in \cite{Tower}, an Heaviside step function can be a good approximation to the Gaussian distribution and will therefore be used in this study.
The only parameter of the distribution will therefore be $\sigma$, which is the higher mass considered in the distribution, so that all scalar fields with $m<\sigma$ are included in the set with the same weight (with the energy density of each scalar field depending only on the initial conditions).

With the set of scalar fields defined, it must now be taken into account that there are two different sectors present in the CTSF, one in which the scalar fields are frozen and the other in which the scalar fields are oscillating, with the mass that separates both sectors being given by $m_s=1/t$, which will be simpler than considering $m_s=H$, as done in \cite{Tower}, for the aimed calculations.

Given that the frozen scalar fields will behave as DE, while the oscillating scalar fields will behave as DM, $\rho_T$ for the CTSF can be obtained by summing the energy for the two mentioned sectors:

\be
\rho_T=\rho_{DM}+\rho_{DE}
\ee

As seen before, $\rho_{DE}$ does not depend on whether the CTSF is in radiation or matter domination. 
Assuming that $t_{ini}=1/\sigma$, $\rho_{DE}$ is given by the following integration of $\rho_{\phi_m}$ over the scalar fields:

\be
\rho_{DE}=C_*\int_0^{m_s} m^{2(1-b)}dm={C_* \over 3-2b}t^{-(3-2b)}
\label{RDE}
\ee

\noindent with $C_*=C/\sigma={1 \over 2}A^2\sigma^{2b-1}$, in order to normalize the energy density.
The energy lost by this sector of the CTSF is only due to the most massive scalar fields start oscillating, when $t=1/m$, and therefore passing to the DM sector.

Unlike the evolution of $\rho_{DE}$, the evolution of $\rho_{DM}$ will depend on the era considered, as the decay in the oscillations will depend on whether the tower is evolving during radiation or matter domination.

Also, it must taken into account that the scalar fields that started oscillating during radiation domination will have a slight different behaviour during matter domination than the ones that start oscillating during matter domination.

For radiation domination, $\rho_{DM}$ is simply given by:

\be
\rho_{DM}=C_*\int_{m_s}^{\sigma} m^{1/2-2b}t^{-3/2}dm={2C_* \over 3-4b}\left(\sigma^{3/2-2b}t^{-3/2}-t^{-(3-2b)}\right)
\ee

\noindent while for matter domination, two distinct cases must be taken into account depending on the value of $\sigma$.

If $\sigma<1/t_e$, the evolution of $\rho_{DM}$ is described by:

\be
\rho_{DM}=C_*\int_{m_s}^{\sigma} m^{-2b}t^{-2}dm={C_* \over 1-2b}\left(\sigma^{1-2b}t^{-2}-t^{-(3-2b)}\right)
\label{DM<t}
\ee

\noindent while for $\sigma>1/t_e$, it must be taken into account that some of the scalar fields in the CTSF have started oscillating during radiation domination, which means the integral to determine $\rho_{DM}$ must be split as:

\ba
\rho_{DM}=C_*\times\left\{
\begin{array}{c l}      
    \int_{m_s}^{m_e} m^{-2b}t^{-2}dm={1 \over 1-2b}\left(m_e^{1-2b}t^{-2}-t^{-(3-2b)}\right)\\
    +\\
    \int_{m_e}^{\sigma} \sqrt{t_e}m^{1/2-2b}t^{-2}dm={2 \over 3-4b}\left(\sigma^{3/2-2b}-m_e^{3/2-2b}\right){t^{-2} \over \sqrt{m_e}}\\
\end{array}\right.
\ea

\be
\rho_{DM}={2C_* \over 3-4b}{\sigma^{3/2-2b} \over \sqrt{m_e}}t^{-2}+{C_* \over (1-2b)(3-4b)}m_e^{1-2b}t^{-2}-{C_* \over 1-2b}t^{-(3-2b)}
\ee

\be
\rho_{DM}={C_* \over 3-4b}\left(2{\sigma^{3/2-2b} \over \sqrt{m_e}}+{1 \over 1-2b}m_e^{1-2b}\right)t^{-2}-{C_* \over 1-2b}t^{-(3-2b)}
\label{DM>t}
\ee

\noindent with $m_e=1/t_e$.

~

\begin{figure}[ht]
\includegraphics[width=8cm]{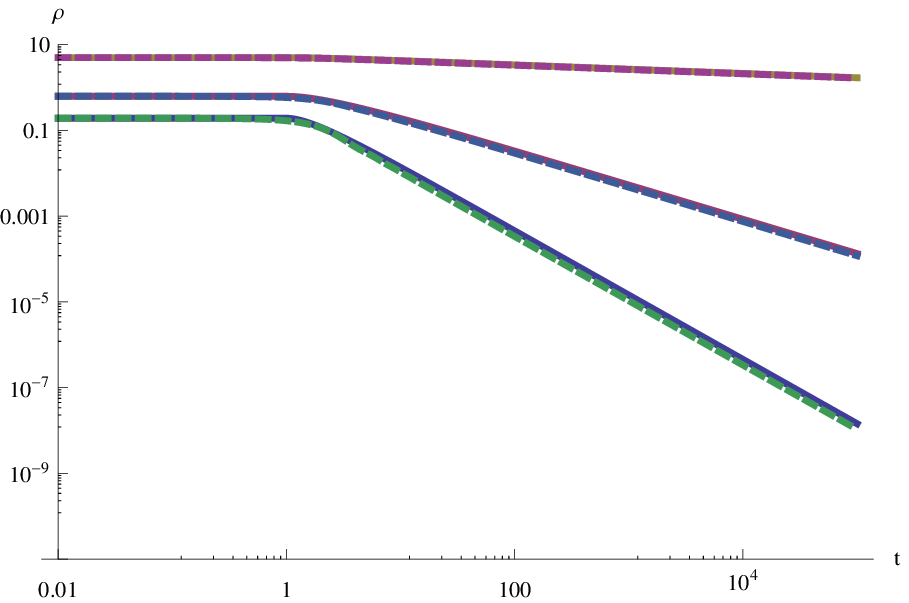}
\includegraphics[width=8cm]{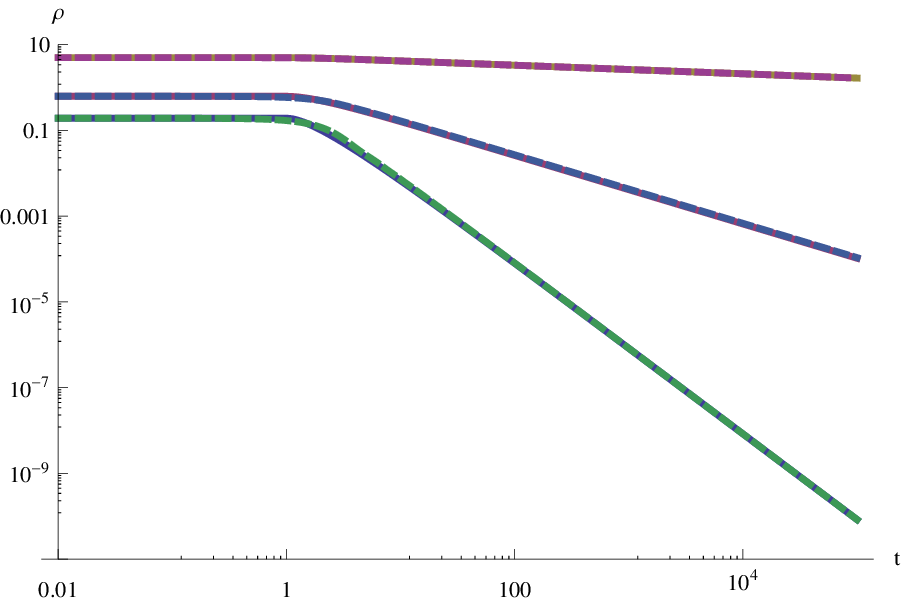}
\caption{
In the left panel, the evolution of $\rho_T$ is displayed for 3 different sets of $b$, the initial parameter, while the right panel shows the same quantities during matter domination.
The values of $b$ considered in the plots are the following, $b=0.2$, $b=1.1$ and $b=1.45$, with the decay of $\rho_T$ slowing as the value of $b$ is raised.
In both plots, the thick line denotes the approximation considered in this section, while the dashed lined refers to the analytic solution obtained in \cite{Tower}.
} 
\label{TRho}
\end{figure}

In order to understand how well the previously obtained solutions describe the evolution of the CTSF, one can compare the results obtained with the integration of the analytic solutions presented in \cite{Tower}, considering the same set of scalar fields in both cases, which leads to figure \ref{TRho}.

\begin{figure}[ht]
\includegraphics[width=8cm]{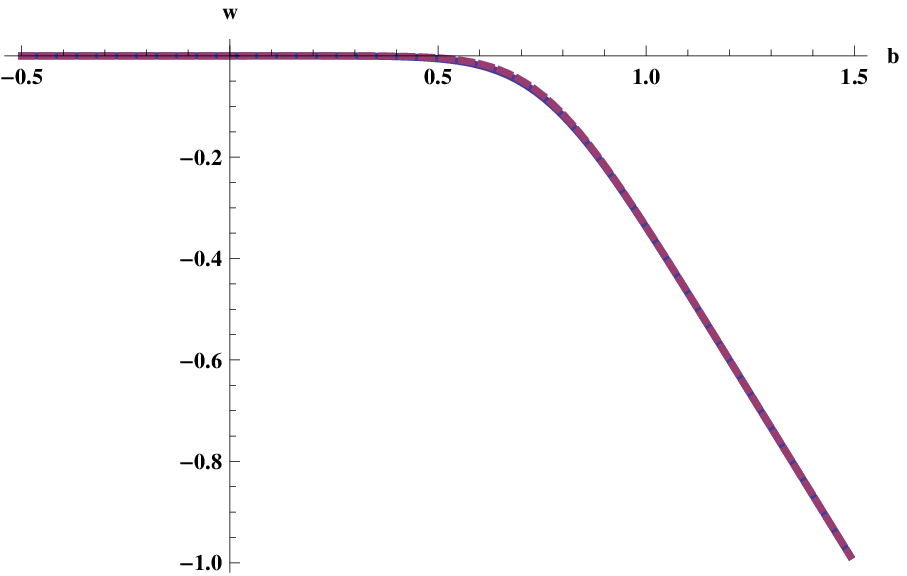}
\includegraphics[width=8cm]{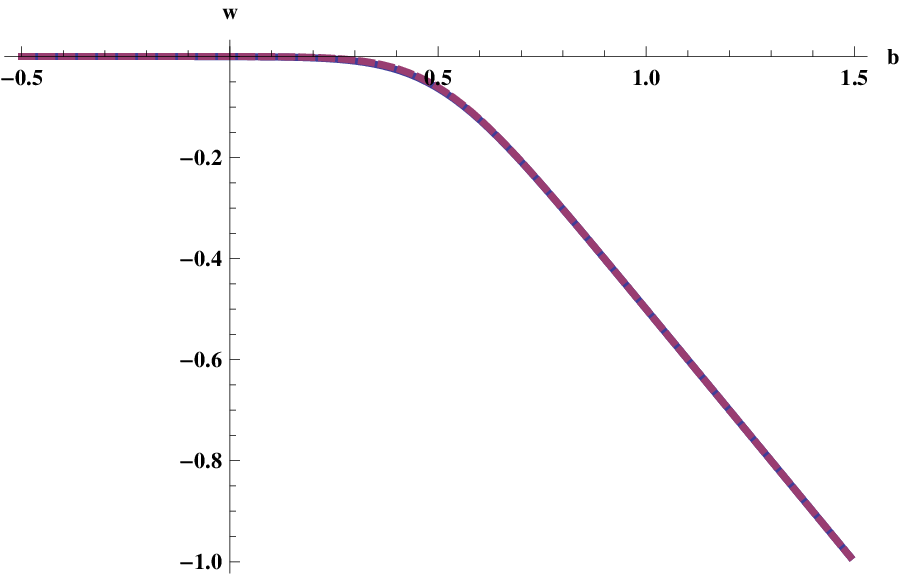}
\caption{
In the left panel it is shown how the equation of state for the CTSF depends on the initial parameter $b$, $w(b)$, during radiation domination, while the right panel shows the same quantity during matter domination. 
In both plots, the thick line denotes the approximation considered in this section, while the dashed lined refers to the analytic solution obtained in \cite{Tower}.
} 
\label{Wb}
\end{figure}

If one aims to take this comparison further, it can be useful to compute how the equation of state, $w_T$, changes with the initial parameter $b$ in both described cases, which leads to figure \ref{Wb}.
This parameter is in fact a way to parameterise the growth or decay of $\rho_T$, given that $\rho_T\propto a^{-3(1+w_T)}$, and therefore, by comparing the quantities for both cases, it is possible to understand how similarly $\rho_T$ varies in both cases.

Again, in both cases, only small differences are observed, meaning the simplified solutions are indeed a good match to the analytic solutions for the CTSF as well.

~

It must be mentioned that the previous results for $\rho_T$ are only valid for $t_{ini}=1/\sigma$.

If $t_{ini}<1/\sigma$, all the scalar fields in the CTSF will remain frozen until $t=1/\sigma$, and hence behaving like DE.
This means $\rho_{DE}={C_* \over 3-2b}\sigma^{3-2b}$ and $\rho_{DM}=0$ for $t<1/\sigma$, considering the chosen set of scalar fields.

On the other hand, if $t_{ini}>1/\sigma$, all the scalar fields with masses $\sigma<m<m_i=1/t_{ini}$ will start oscillating at $t_{ini}$. This means there will be an extra amount of DM that must be taken into account, which will only affect the evolution of $\rho_{DM}$.

The new evolution for $\rho_{DM}$ can be obtained by replacing $\sigma$ in the expressions for $\rho_{DM}$, presented in eq.\ref{DM<t} and \ref{DM>t}, with $m_i$ and then adding the following term:

\be
\rho_{DM_E}=\left({t \over t_{ini}}\right)^{-3p}C_*\int_{m_i}^{\sigma} m^{1/2-2b}t^{-3/2}dm={C_* \over 3-2b}\left({\sigma^{3-2b}-m_i^{3-2b}}\right)\left({t \over t_{ini}}\right)^{-3p}
\ee

\noindent which corresponds to the energy of the extra amount of DM, taking into account it decays with $a^{-3}$.

~

Also, the results for $w(b)$ presented in figure \ref{Wb} will not depend on the set of scalar fields considered for the CTSF, given those are asymptotic results.

However, extending the width of the distribution of the CTSF, in order to add an extra set of scalar fields with higher masses, can have a significant impact in the value of $w(b)$ at a given time.

In fact, if $b$ is large, the only way the CTSF can account for DM as well as DE is to consider the previously mentioned extension, so that the DM sector will include more scalar fields.
Given that extra set will have a significant contribution to $\rho_T$, it is possible to obtain $w(b)\approx0$ even for high values of $b$, as mentioned in \cite{Tower}.  

~

If the CTSF is aimed to account for both DM and DE, the expression obtained for $\rho_T$ in \cite{Tower} is actually better than the ones presented in this paper.

As can be seen, the presented expressions for $\rho_{DM}$ and $\rho_{DE}$ both present singularities, depending on the value of $b$ considered.
The singularity presented in the expression for $\rho_{DE}$, corresponding to $b\rightarrow3/2$, is expected and was mentioned in the discussion on \cite{Tower}, but the singularities presented in the expression for $\rho_{DM}$, considering $b\rightarrow1/2$ and $b\rightarrow3/4$, are not physical and appear due to the approximation considered. 

However, the expressions determined in this paper are actually better when considering that the CTSF does not have to account for DM, given that it does not rely in the assumption that the DM sector of the CTSF dominates at $t_{ini}$.

~

As a final note, despite $m_s=1/t$ in this paper rather than $m_s=H$, as used in \cite{Tower}, the cosmological results obtained previously in \cite{Tower} shall hold for this paper and the parameters shall be of the same order of magnitude with the present consideration for $m_s$.
Also, this new choice for $m_s$ is solely due to it being better to simplify the calculations presented in this paper, but still a reasonable choice, as seen in the results presented in figure \ref{1Field}.

\section{Dark Matter - Dark Energy Interaction in a Continuous Tower of Scalar Fields}

As seen in the previous section, the CTSF can easily be split in two sectors, one that behaves like DE while the other behaves like DM.
The mentioned splitting is possible due to the simplification considered, as it forces the scalar fields in the CTSF to be either in a DE or DM state, with the transition from a DE behaviour to a DM behaviour occurring instantaneously at $t=1/m$.

Since the scalar fields change from a DE state, when they are frozen, to a DM state, when they oscillate, the scalar fields are in fact transferred from the DE sector of the CTSF to the DM sector at $t=1/m$.
Therefore the transference of energy from one sector to the other can be seen as an exchange of scalar fields from the DE sector to the DM sector and hence the energy that is transferred at a given time will depend on the energy possessed by the scalar fields that change states.
This is considerably different than the usual interacting DM-DE models, in which the energy transferred at a given time depends in the energy of one or both of the whole fluids involved in the interaction.

The energy transferred is given by the coupling $Q$, which will therefore depend in the energy of the scalar fields that start oscillating at a given moment.
Said scalar fields will cover a small range in the mass spectrum, with $m \approx m_s=1/t$, whose energy is given by eq.\ref{inen}.

~

Taking into account the normalisation due to the continuous distribution of scalar fields, $Q$ can be written as:

\be
Q={1 \over \sigma}\rho_{\phi_{m_s}}(t_0){dm_s \over dt}=-C_*t^{-(4-2b)}
\ee

\noindent assuming that the CTSF is evolving during radiation or matter domination.

~

Therefore, one finds $Q\propto a^{-(2-b)}$ during radiation domination, and $Q\propto a^{-{4 \over 3}(2-b)}$ during matter domination, given that $a\propto t^{p}$, as mentioned before.

Switching back to time, the equations that take into account the interaction between the two components are given by:

\ba
\begin{array}{c l}      
    \dot{\rho}_{DE}=-C_*t^{-(4-2b)}~,\\
    ~\\
    \dot{\rho}_{DM}+3H\rho_{DM}=C_*t^{-(4-2b)}~,\\
\end{array}
\ea

\noindent assuming that the scalar fields in the DE sector are frozen, which means $w_{DE}=-1$, and taking into account that the energy flows from the DE sector into the DM sector.

~

As can be seen clearly, both equations are not actually coupled, which means it is possible to obtain the evolution for $\rho_{DM}$ and $\rho_{DE}$ independently.

This is one of the biggest advantages of this kind of parameterisation, given that this is not the case with conventional models that consider $Q$ to be parameterised by $Q=\delta_{DE}H\rho_{DE}+\delta_{DM}H\rho_{DM}$, for which a more complex dynamical systems analysis must be considered in order to understand how $\rho_{DE}$ and $\rho_{DM}$ evolve.

~

Starting with the equation for DE, since it does not depend on the era considered, the following result is obtained:

\be
\rho_{DE}={C_* \over 3-2b}t^{-(3-2b)}
\ee

\noindent which fully agrees with eq.\ref{RDE} obtained in the previous section.

To obtain $\rho_{DM}$, the era in which the CTSF is evolving must be again taken into account, as the evolution of $H$ will depend on the era considered.
Therefore, for radiation domination $\rho_{DM}$ will evolve as:

\be
\rho_{DM}=C_1t^{-3/2}-{2C_* \over 3-4b}t^{-(3-2b)}
\ee

\noindent while for matter domination the evolution will be given by:

\be
\rho_{DM}=C_2t^{-2}-{C_* \over 1-2b}t^{-(3-2b)}
\ee

~

The previous results obtained for $\rho_{DM}$ are in agreement with eqs.\ref{DM<t} and \ref{DM>t} obtained in the previous section, considering the simplified approximated evolution of the scalar fields, as long as:

\be
C_1={2C_* \over 3-4b}\sigma^{3/2-2b}
\label{C1}
\ee

~

\noindent and:

\ba
C_2=\left\{
\begin{array}{c l}      
    {C_* \over 1-2b}\sigma^{1-2b} & ~,~\sigma<1/t_e\\
    ~\\
    {C_* \over 3-4b}\left(2{\sigma^{3/2-2b} \over \sqrt{m_e}}+{1 \over 1-2b}m_e^{1-2b}\right) & ~,~\sigma>1/t_e\\
\end{array}\right.
\label{C2}
\ea

This means in fact the CTSF can be described as a model of interacting DM-DE during radiation and matter domination, as long as the adequate parameterisation for $Q$ is considered.

~

\begin{figure}[ht]
\includegraphics[width=8cm]{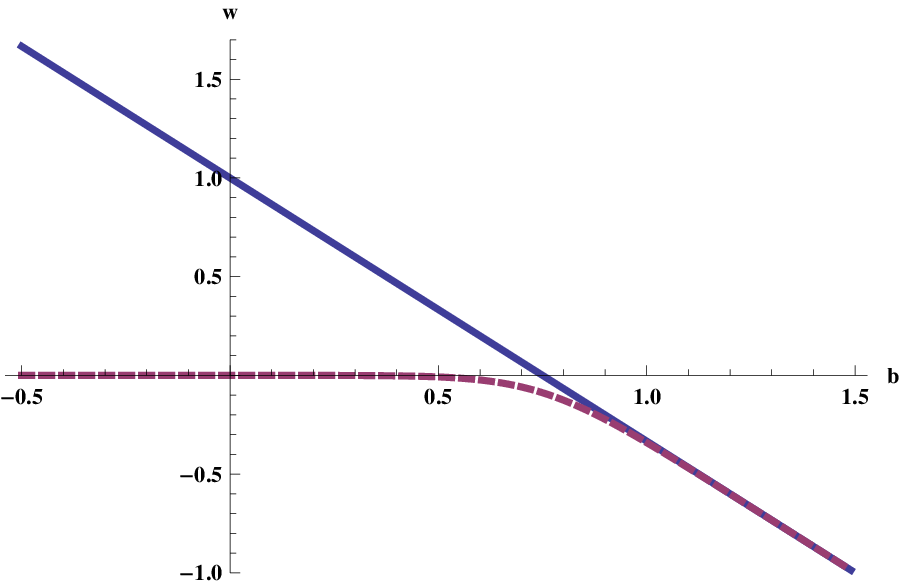}
\includegraphics[width=8cm]{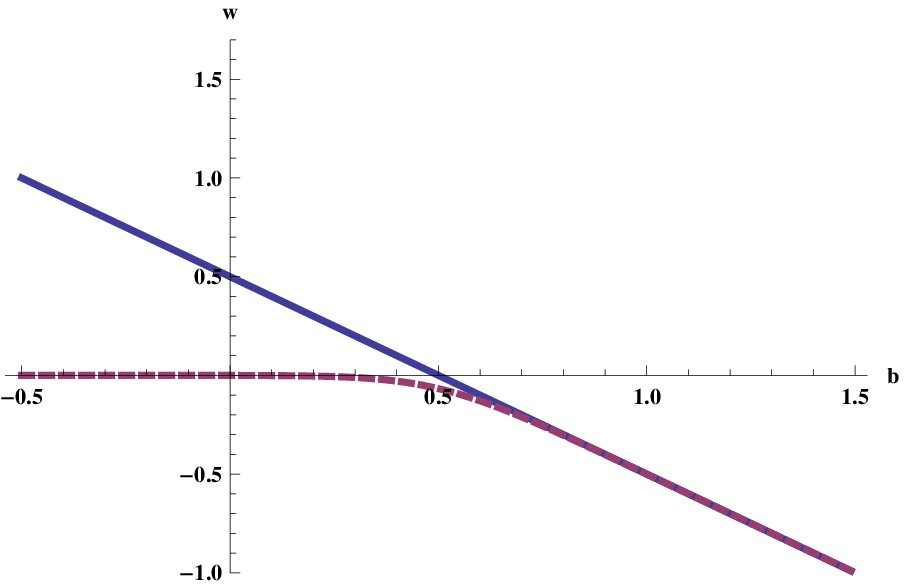}
\caption{
In the left panel, the equation of state for the two sectors of the CTSF, the DM and DE sectors, is plotted in function of the initial parameter $b$, $w(b)$, during radiation domination, while the right plot shows the same quantities during matter domination. 
In both plots, the thick line denotes the DE sector, while the dashed lined refers to the DM sector of the CTSF.
} 
\label{WbS}
\end{figure}

However, it must be pointed out that the effective value for $w_{DM}$ and $w_{DE}$ can be quite far from the expected values ($w_{DM}=0$ and $w_{DE}=-1$), as can be seen in figure \ref{WbS}.

Despite figure \ref{WbS} shows a wide range of possible values for $w_{DM}$ and $w_{DE}$, the scalar fields that compose each of the sectors will behave either as DM or DE at an individual level, which explains the name given to the different sectors.

~

If one takes into account the equivalence to the CTSF shown and use the constraints on the initial conditions obtained in \cite{Tower}, it is easy to understand which values are in fact reasonable for $w_{DM}$ and $w_{DE}$.

As seen in \cite{Tower}, for the CTSF to be in agreement with the observations, the value of $b$ must be close to 1.5, which implies $w_{DE}\approx-1$, but also $w_{DM}\approx-1$, if one considers a distribution for the CTSF with $t_{ini}=1/\sigma$.
Said assumption was used to determine the constants $C_1$ and $C_2$, presented in eqs. \ref{C1} and \ref{C2} respectively, but unfortunately does not allow for the whole system to account for the whole dark sector.

If one wants for the whole system to account for the whole dark sector, an extra amount of DM is required, which will prevent the asymptotic behaviour presented in figure \ref{WbS} to be reached until now and keep $w_{DM}\approx0$.
Said extra amount of DM can be obtained by raising the constants $C_1$ and $C_2$, which in fact corresponds to consider a larger width to the distribution in the CTSF.

~

It must also be pointed out that, given $Q$ does not depend on $\rho_{DM}$, a change in $C_1$ and $C_2$ will not affect the evolution of $\rho_{DM}$ or $\rho_{DE}$, unlike what happens in other models of interacting DM-DE described in the first section.

~

It is worth mentioning that this model does not allow for further constraints to be applied to the parameters on the CTSF model.

Unlike the models mentioned in the introduction, calculating the ratio $R=\rho_{DE}/\rho_{DM}$ does not make sense in this model, given that it breaks down when the DE sector of the CTSF starts dominate.
This way, it is not possible to obtain constraints to the interaction, that would necessary lead to constraints on the parameters of the CTSF model.

Therefore, it is also not possible to use this model to solve the "coincidence" problem due to the same reason, which was expected before hand, given that the asymptotic behaviour of the CTSF is equivalent to a cosmological constant, as found in \cite{Tower}, which means an interacting model using the CTSF could never solve said problem.

~

As a final note, it is relevant to mention that it would have not been possible to establish this equivalence using the expressions presented in \cite{Tower} for the different $\rho$'s, given that it was assumed that the whole energy of the CTSF is on the DM sector at $t_{ini}$.
Despite this is a good approximation when the CTSF accounts for both DM and DE, this approximation prevents any energy from being transferred to the DM sector, which is crucial to establish this equivalence.

\section{Conclusions}

As aimed, this paper presents a new parameterisation for $Q$, the coupling between DM and DE, physically supported by an equivalence between the CTSF, first introduced in \cite{Tower}, and a system of interacting DM-DE.

This analogy is only possible if the scalar fields belonging to the CTSF are assumed to change from a DE behaviour to a DM behaviour instantaneously.
As seen before, this is only reasonable for radiation and matter domination, but not when the DE sector of the CTSF is the dominant component of the universe.
Despite this drawback, the results obtained for the CTSF considering this approximation for the scalar fields pretty much match the results obtained using the analytic solutions for the scalar fields presented in \cite{Tower}.

By splitting the CTSF in a DM and a DE section and treat it as a DM-DE interacting system, it is possible to obtain the same results as before, as long as it is considered that the DM fluid has $w=0$, while $w=-1$ for the DE fluid, and also that $Q$ corresponds to the energy of the scalar fields that start oscillating at a given time and hence change from a DE state into a DM state.

This way it is clear the CTSF is basically equivalent to a system of interacting DM-DE during radiation and matter domination, which introduces a new perspective to how the CTSF actually works.

Furthermore, this equivalence allows the introduction of a different parameterisation for $Q$ that does not depend on any properties of the DM or DE fluids, but rather on the energy of the individual scalar fields that constitute the DM and DE fluids. 
The energy is therefore quantised in the scalar fields that constitute the CTSF, giving a whole new perspective on how the interactions between DM and DE can work.

~

Despite this paper focus only on the CTSF model and how it can be seen as an interacting DM-DE model, it is possible to generalise the calculations done to fit other models involving multiple scalar fields.

Once again, it is important that the transitions a scalar field undergoes from one state to another occur quickly, so that it is possible to approximate said transition to be instantaneous if the interacting fluid approach is to be applied to other systems with multiple scalar fields.
However, it is not necessary that the scalar field can only be in a DM or DE states for this formalism to be applied, as long as it is taken into account the actual value of $w$ for each of the state the scalar field can be at and that the expression for $Q$ is modified so that it corresponds to the actual energy a scalar field carries when switching states.
Given that $Q$ will not depend on the total energy of DM or DE, this also means it is much easier to understand the evolution of $\rho_{DM}$ and $\rho_{DE}$ without resorting to a dynamical systems analysis.

It is also relevant to point out that the mass in the multiple scalar fields system does not have to be continuous.
However, a discrete distribution over the mass yields discrete transferences in the energy from one component to the other.
This would be relevant for the model presented in \cite{TDM}, for instance.

~

Last but not least, it is also relevant to mention that the approximations introduced for the evolution of $\rho_{\phi_m}$ can be quite useful, given that it allows for a simple and accurate expression to be obtained for $\rho_T$, as seen in section 2.
Despite the expressions obtained end up not being as simple as the ones obtained in \cite{Tower}, these expressions are valid when the DM sector of the CTSF considered is not clearly dominant at $t=t_0$, which is not the case with the expressions presented in \cite{Tower}, given those rely on $\rho_{DM}\approx\rho_T$ at the initial time.
Therefore, any future project featuring the CTSF can benefit from this simplification as it will speed up significantly any computational calculations that are required for radiation and matter domination, given that it is possible to use a simple analytic expression rather than a considering the numerical integration of analytic solutions for $\phi_m$.

\section*{Acknowledgments}

The work of P.S. is sponsored by FCT - Funda\c{c}\~ao para a Ci\^encia e Tecnologia, under the grant SFRH / BD / 62075 / 2009. P.S. is very thankful to Dr. Jo\~ao Rosa, Dr. Jose Beltr\'an Jim\'enez, Dr. Nelson Nunes and to everyone who attended Cosmonata 2014 for the interesting discussions that allowed to improve the paper and also to Prof. David Mota and Prof. \O ystein Elgar\o y for the motivation given.

\section*{References}


\end{document}